\def\year{2017}\relax
\documentclass[letterpaper]{article}
\usepackage{aaai18}
\usepackage{times}
\usepackage{helvet}
\usepackage{courier}
\usepackage{url}
\usepackage{graphicx}
\usepackage{subcaption}
\usepackage{amssymb,amsmath}
\usepackage{url}

\usepackage{color}
\usepackage{ifthen}
\usepackage{newclude}
\usepackage{epstopdf}
\usepackage{flushend}
\usepackage{xspace}
\usepackage{graphicx}
\usepackage{tikz}
\usepackage{pgfplots, pgfplotstable}
\usepgfplotslibrary{statistics}
\usepackage{xspace}
\usepackage{wasysym}
\usepackage{eurosym}

\makeatletter
\def\url@foostyle{%
  \@ifundefined{selectfont}{\def\UrlFont{\rm}}{\def\UrlFont{\rmfamily}}}
\makeatother
\usepackage[hyphenbreaks]{breakurl}
\usepackage[hang,flushmargin]{footmisc}

\begin{document}

\sloppy

\title{Large Scale Crowdsourcing and Characterization of Twitter Abusive Behavior}

\author{Antigoni-Maria Founta$^\ddagger$,
Constantinos Djouvas$^\star$,
Despoina Chatzakou$^\ddagger$,
Ilias Leontiadis$^\sharp$,\\
{\Large \bf Jeremy Blackburn$^\mathsection$,
Gianluca Stringhini$^\dagger$,
Athena Vakali$^\ddagger$,
Michael Sirivianos$^\star$,
Nicolas Kourtellis$^\sharp$}\\
$^\ddagger$Aristotle University of Thessaloniki \hspace{0.2cm}
$^\star$Cyprus University of Technology \hspace{0.2cm}
$^\sharp$Telefonica Research \hspace{0.2cm}\\
$^\mathsection$University of Alabama at Birmingham \hspace{0.2cm}
$^\dagger$University College London\\
{\{founanti,deppych,avakali\}@csd.auth.gr}, \{costas.tziouvas,michael.sirivianos\}@cut.ac.cy, \\
{\{nicolas.kourtellis,ilias.leontiadis\}@telefonica.com}, {blackburn@uab.edu},{g.stringhini@ucl.ac.uk}\\}

\maketitle

\begin{abstract}
In recent years online social networks have suffered an increase in sexism, racism, and other types of aggressive and cyberbullying behavior, often manifesting itself through offensive, abusive, or hateful language.
Past scientific work focused on studying these forms of abusive activity in popular online social networks, such as Facebook and Twitter.
Building on such work, we present an eight month study of the various forms of abusive behavior on Twitter, in a holistic fashion. Departing from past work, we examine a wide variety of labeling schemes, which cover different forms of abusive behavior. We propose an incremental and iterative methodology that leverages the power of crowdsourcing to annotate a large collection of tweets with a set of abuse-related labels. By applying our methodology and performing statistical analysis for label merging or elimination, we identify a reduced but robust set of labels to characterize abuse-related tweets. Finally, we offer a characterization of our annotated dataset of 80 thousand tweets, which we make publicly available for further scientific exploration.
\end{abstract}

\section{Introduction}\label{sec:intro}
The rise of hateful behavior online has recently become a topic of interest. The research community has studied hate speech~\cite{djuric2015hate}, cyberbullying~\cite{hosseinmardi2015analyzing}, and semi-organized online harassment campaigns~\cite{hine2017kek}, while also proposing systems to automatically detect and block abusive behavior~\cite{ribeiro2017like,davidson2017automated,serra2017class}. To their credit, social network platforms are also taking steps to mitigate damage, e.g., providing users with tools to flag abusive behavior~\cite{kayes2015ya-abuse,twitter2017}. 

Unfortunately, abusive content poses some unique challenges to researchers and practitioners. First and foremost, even defining what qualifies as abuse is not straightforward, which in turn makes it difficult to extract ground truth to base deeper exploration off of. Unlike other types of malicious activity, e.g., spam or malware, the accounts carrying out this type of behavior are usually \emph{controlled by humans}, not bots. This makes techniques based on grouping together similar messages or searching for automated activity ineffective. Second, this activity is not particularly common; relatively few examples can be found from a random collection of posts.
Elaborate techniques must be employed to boost the collection of minority examples for the automated techniques to work.

To deal with these issues, crowdsourcing proved to be a promising direction towards developing labeled datasets. However, human labeling poses a number of challenges. One of the main issues is the existence of different types of abuse and different labels describing them (e.g., offensive language, hate speech, aggressive behavior, cyberbullying, etc.). It is often difficult, even for a human, to consistently distinguish them~\cite{chatzakou2017mean}. For example, certain types of language, such as sarcasm, can be misinterpreted by annotators if the messages are not presented in context. Another challenge is striking a good balance between the number of annotators employed per task, their payment, and how much time the crowdsourcing process takes to complete.

This paper tackles three challenges faced when trying to collect large scale ground truth on abusive behavior: (i) difficulty of the crowdsourced workers to distinguish between different abusive categories (e.g., hate speech vs. offensive language vs. abusive language), (ii) different occurrence rates for different categories of abuse, and (iii) scaling the multi-labeled annotation process to thousand tweets, while maintaining quality of annotation and time-budget constraints.

Into the direction of addressing the aforementioned challenges we proceed with the following contributions:
\begin{itemize}
	\item A methodology to detect and cut through the confusion of crowdworkers when they are asked to distinguish between nuanced labels.
	\item A boosted sampling approach that maintains an unbiased dataset, while ensuring more annotations for the minority class.
	\item The design and development of a data collection platform that optimizes for costs.
	\item A dataset of 80k tweets, annotated for abusive behavior and created following the methodology and platform we developed and present in this paper.\footnote{The dataset is hosted at: \url{http://ow.ly/BqCf30jqffN}.}
	\item The open source code of the platform used for collecting crowdsourced annotations from users.\footnote{The code can be found at: \url{http://ow.ly/TnuU30jqf7g}.}
\end{itemize}

\section{Related Work}\label{sec:back}

\begin{table*}[hpt!]
	\centering
	\small
	\begin{tabular}{|c|c|c|c|}
		\hline
		\bf{Dataset}						&\bf{\# Tweets}	&	\bf{Labels}								&\bf{Annotators}	\\ \hline
		\cite{chatzakou2017mean}			&	$9,484$		&	aggressive, bullying, spam, normal				&	5			\\ \hline
		\cite{waseem2016hateful}				&	$16,914$	&	racist, sexist, normal							&	1			\\ \hline
		\cite{davidson2017automated}			&	$24,802$	&	hateful, offensive (but not hateful), neither			&	3 or more		\\ \hline
		\cite{golbeck2017harassment-dataset}	&	$35,000$	&	the worst, threats, hate speech, direct harassment,	&	2-3			\\
		&			&	potentially offensive, non-harassment			&				\\ \hline
		Present study						&	$80,000$	&	offensive, abusive, hateful speech, aggressive,		&	5-20			\\
		&			&	cyberbullying, spam, normal					&				\\ \hline
	\end{tabular}
	\caption{Summary of related work datasets.}
	\label{tab:rel-work-datasets}
\end{table*}

In the recent past, a few datasets have been collected, annotated, and released by other researchers around abusive behavior on Twitter, using various labels and different methodologies for annotation.

Wasseem and Hovy~\cite{waseem2016hateful} provide a dataset used by several studies working on hate speech detection~\cite{badjatiya2017deep,gamback2017using,jha2017does}. That work focused on disambiguating different types of hate speech, and more specifically between \emph{racism} and \emph{sexism}, at the level of tweets (i.e., whether a tweet is racist or sexist). The authors collected tweets based on a set of hate-related terms and users, and manually annotated a subset of their dataset using an outside annotator for reviewing. The final dataset consists of almost $17k$ tweets, $12\%$ of which are labeled as racist, $20\%$ sexist, and the rest are considered normal. After the annotation process, they investigated which features better assist the detection of hate speech by building upon this dataset.
They show that only gender plays an important role, while geographic and word-length distributions are typically ineffective.

\cite{davidson2017automated} provide a dataset that has also been studied, e.g.,~\cite{malmasi2017detecting,olteanu2017limits}. The focus of this work was mainly to distinguish between \emph{hateful} and \emph{offensive} language. According to the authors, offensive language contains offensive terms which are not necessarily inappropriate, while hate speech intends to be derogatory, humiliating, or insulting. Focusing on the differences between the two, they started their annotation process by identifying and collecting a set of possible hateful users and extracted their tweets. Then, they sampled this collection for $25k$ tweets containing terms from a hate speech lexicon. Finally, this sampled dataset was annotated by CrowdFlower workers. With an intercoder-agreement score of $92\%$, the majority label was offensive ($77\%$), a small percentage hateful ($6\%$), and the rest were normal.

\cite{golbeck2017harassment-dataset} focused on online trolling and harassment on Twitter. They first used various online sources, e.g., blocklists, to produce a list of keywords used for collecting harassing tweet with high probability. Subsequently, they created guidelines for the annotation task and trained coders to label the tweets using labels such as ``the very worst,'' ``threats,'' ``hate speech,'' ``direct harassment,'' ``potentially offensive,'' and ``non-harassment.'' Their aim was to label tweets as harassing only if they really were ``the worst of the worst content.'' Their final dataset includes $35k$ tweets annotated by 2 or 3 coders.

The previous datasets focus on the language used by annotating text content. However, there are other works focusing on user characteristics, i.e., they provide annotation of Twitter users based on their exhibited behavior.
As \cite{ribeiro2017like} emphasize, identifying content as hateful raises major issues, while ``characterizing and detecting hateful users [...] presents plenty of opportunities to explore a richer feature space.'' Therefore, detecting inappropriate user behavior is a different, but closely related task. \cite{chatzakou2017mean} detected cyberbullying and cyber-aggression by collecting and annotating a dataset from Twitter. Their annotation methodology is similar to ours in the use of CrowdFlower for the annotation task. Their final dataset contains $9,484$ tweets, labeled as one of four categories: 1)~bullying, 2)~aggressive, 3)~spam, or 4)~normal. The aggressive and bullying labels make up about $8\%$ of the dataset, spam makes up about $33\%$, and the remainder of the annotations are normal.

While the previous works fall under the same domain, i.e., annotating inappropriate speech, one crucial challenge still remains unaddressed. Specifically, there is an important gap regarding the principled selection of the most appropriate labels for annotating aggressive online behavior. In past literature, types or labels of inappropriate speech are usually used interchangeably, or selected randomly among available ones for use in the annotation task. Indeed, past studies do not explain or justify the selection of types of inappropriate speech which they use for their annotations. Furthermore, they either use only a subset of popular labels, and consider the others covered (without however establishing why this is so), or combine them together under the same umbrella label. In this work, we take a step back and propose a principled methodology to narrow down the list of possible labels used in this space. Our methodology is iterative, to account for limited time and budget, as well as allowing for controlled statistical analysis of label selection by annotators. We use the final set of selected labels for a large scale crowdsourcing study, annotating $80k$ tweets with appropriate labels on abusive behavior. Table~\ref{tab:rel-work-datasets} summarizes the past works relevant to the topic which have released their datasets for scientific exploration.

\section{Overview of Methodology}\label{sec:methodology}

\subsubsection{Our goal:}
The overall goal of this work is to create a large and highly accurate annotated dataset of tweets ($80k$) via a crowdsourcing platform like CrowdFlower (\textbf{CF}). Unlike previous work, we are interested in workers selecting from more than one potential category of abusive behavior (i.e., two or more labels), in order to study the correlation between them and make adjustments on the final labels used. Annotating such a large dataset exhibits some unique challenges since we must \emph{minimize the cost without compromising the quality} of the annotation. This can only be achieved by carefully tuning the task and the platform, appropriately selecting the samples to annotate, and striking a balance between worker payment and quality.

\subsection{Challenges with Crowdsourced Platforms}
There are several challenges that need to be resolved to best balance high-quality annotations and minimal cost.

\subsubsection{Labels:}
We want to build a dataset that distinguishes between various expressions of online abuse, e.g., abusive and aggressive, hateful, offensive, cyberbullying. However, even if detailed definitions and examples are provided, crowdsourced workers might still find it challenging to consistently label examples. Thus, the overall design of the task (e.g., how to phrase the questions) and the selection of labels to choose from is an important challenge. To address this, we use a number of preliminary annotation rounds that aim to identify the exact nature of any confusion. This facilitates the elimination of any ambiguity on the labels used during the main annotation and, thus, achieving high accuracy and consistency.

\subsubsection{Sampling:} In the grand scheme of things, abusive tweets are quite rare (between $0.1\%$ and $3\%$, depending on the label).
Therefore, even large-scale datasets might contain just a few samples. For typical machine learning algorithms which can take benefit from such a dataset, few samples means less opportunity to train on the specific behaviors, and overall, worse classification performance. One way to deal with this extreme imbalance is to pre-select tweets that are likely to be abusive (e.g., those that contain known hate words), however, this approach also biases the dataset.

To address this sampling issue, we designed a boosted random sampling technique. A large part of the dataset is randomly sampled, but then boosted with tweets that are likely to belong into one or more of the minority classes. We use text analysis and preliminary crowdsourcing rounds to design a model that can pre-select the tweets of the boosted set. Both sets are then mixed together and given to the crowdsourcing platform for the final annotation.

\subsubsection{Judgments:}
The next challenge to address is determining the proper number of crowdworker decisions that are necessary for a high confidence annotation. As expected, this is largely dependent on a combination of factors: complexity of the task, worker reward, quality of annotators, etc. The solution we settled in is to employ a large number of annotators during the exploratory rounds (up to 20 annotators per tweet) to establish the general level of agreement we should expect, given the number of annotators.

\subsubsection{Payment:}
The payment to crowdworkers also plays an important role in their annotations. Studies have shown that when participants are payed fairly, it positively affects their results, but always depending on the type of task to be performed \cite{ye2017crowdworkers-payment}. In our case, and similarly to~\cite{chatzakou2017mean}, we started with a default payment scheme ($5\cent$ for a batch of 10 tweets), and used the preliminary rounds to adjust as needed.

\subsection{Crowdsourcing Methodology}
Our methodology addresses the above challenges via a three-step process. We visualize these steps in three figures: Figure~\ref{fig:datapipeline} illustrates the preliminary data preparation process. Figures~\ref{fig:exploratoryanalysis} and~\ref{fig:final} visualize the next two steps which involve the iterative annotation rounds.

\subsubsection{Step 1: data collection and sampling.}

\begin{figure}[t]
	\centering
	\includegraphics[width=\columnwidth]{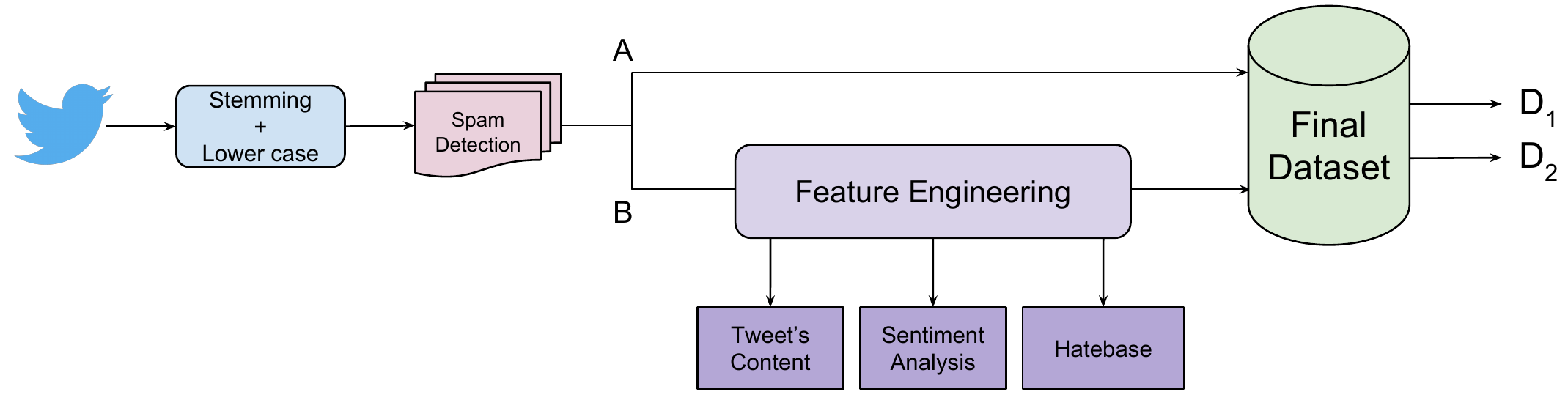}  
	\caption{Data Preparation Pipeline (Step 1). Pre-filtering and spam removal to clean tweets. ($A$) random set of un-boosted tweets. ($B$) boosted sampling to produce a set of tweets biased towards abusive behavior. Sub-datasets $D1$ and $D2$ are used in the subsequent Steps 2 and 3.}
	\label{fig:datapipeline}
	\vspace{-1em}
\end{figure}

The first step of the process (Figure~\ref{fig:datapipeline}) is to collect a random set of tweets. To do so, we utilize the Twitter Stream API. We store the data in elastic search and we apply basic pre-filtering to exclude spam, tweets that have no content, tweets that are not in English, etc. Furthermore, we apply simple text analysis and machine learning to create the boosted set of tweets, that will be used to improve coverage over the minority classes. Finally, we randomly sample a small dataset (D1), that is used for the exploratory analysis, and the remainder (D2) for the large scale annotation.

\subsubsection{Step 2: exploratory analysis.}

\begin{figure}[t]
	\centering
	\includegraphics[width=1\columnwidth]{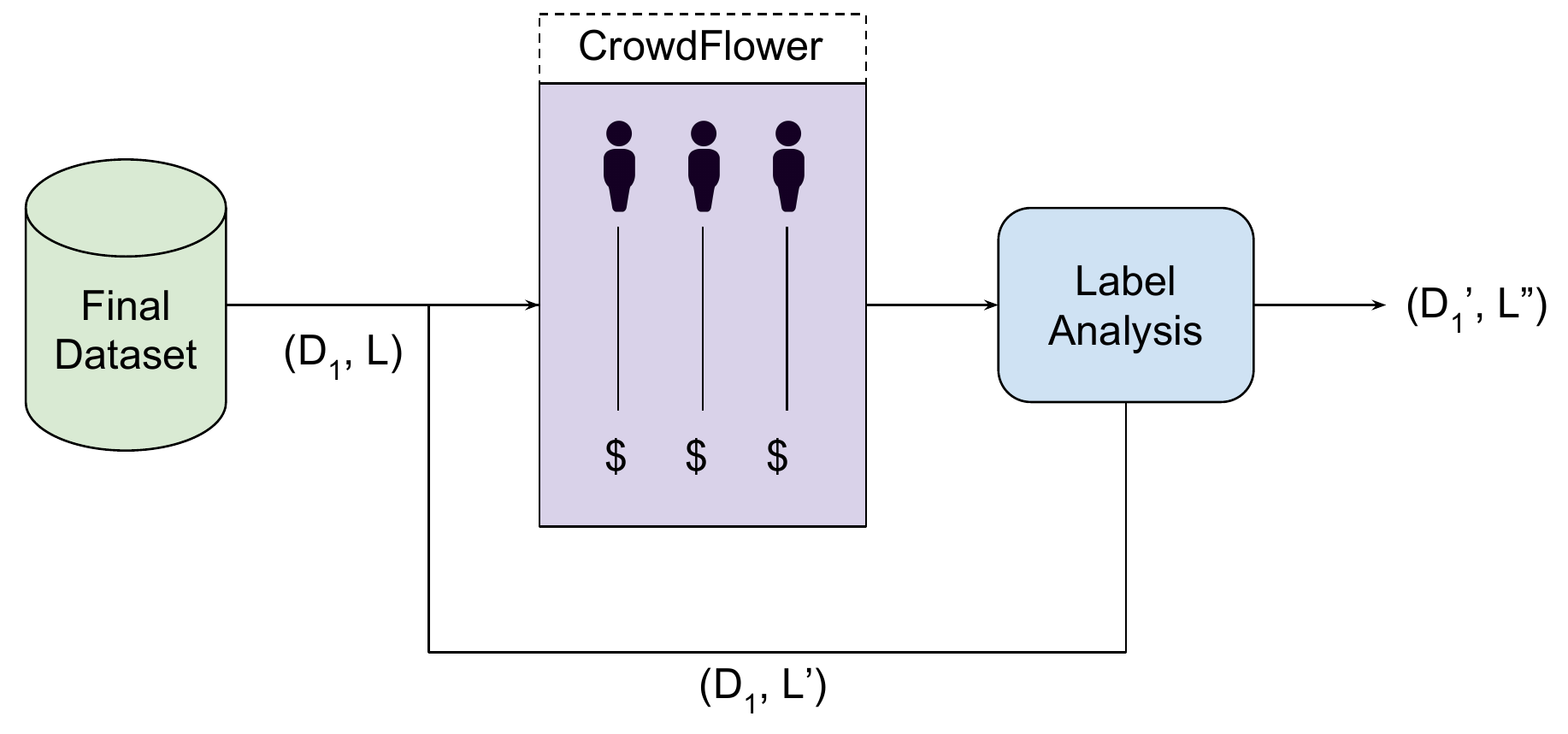}  
	\caption{Exploratory Analysis (Step 2). Dataset $D1$ is inputed in the platform for annotation under label set $L$, and in consecutive rounds. In each round, statistical analysis performed can narrow down the set of labels to $L'$. Final set of labels $L''$ can be inputed in Step 3.}
	\label{fig:exploratoryanalysis}
	\vspace{-1em}
\end{figure}

Considering the various trade-offs among the different parameters of the annotation task, we first introduce an iterative process that allows the researchers to properly adjust these parameters (Figure~\ref{fig:exploratoryanalysis}). This analysis is performed on a small sample ($300$ tweets in our case), as a means to enable quick and affordable testing among all the different design choices and parameters. These parameters include the payment, type and number of labels, presentation of these labels, number of judgments required, trustworthiness of users, annotation process, etc. Furthermore, this process can reveal possible points of confusion (e.g., identify if two labels are frequently mixed).

During these iterations, we fine-tune the filters used to better boost the dataset, in order to contain more samples of the minority classes. After each iteration, an analysis of the results can reveal if a satisfying quality is reached and whether a given parameter has contributed to this, akin to an A/B testing. In an ideal scenario, a researcher can execute many such iterations to optimize better for the set of labels to be used, the money paid to workers, etc. In practice, and always due to limited budget and time, these iterations can only be a handful.

In our case, the process converged after three iterations, allowing us to identify the influence each one of the aforementioned parameters has in the annotated dataset. The main outcome of this step was to determine the most representative and clearly understood set of labels that should be used for the large scale annotation task. It also enabled us to assess the ideal number of judgments to strike a balance between cost and quality. Details about each round are given in the following sections.

\subsubsection{Step 3: full annotation.}

\begin{figure}[tp]
	\centering
	\includegraphics[width=0.9\columnwidth]{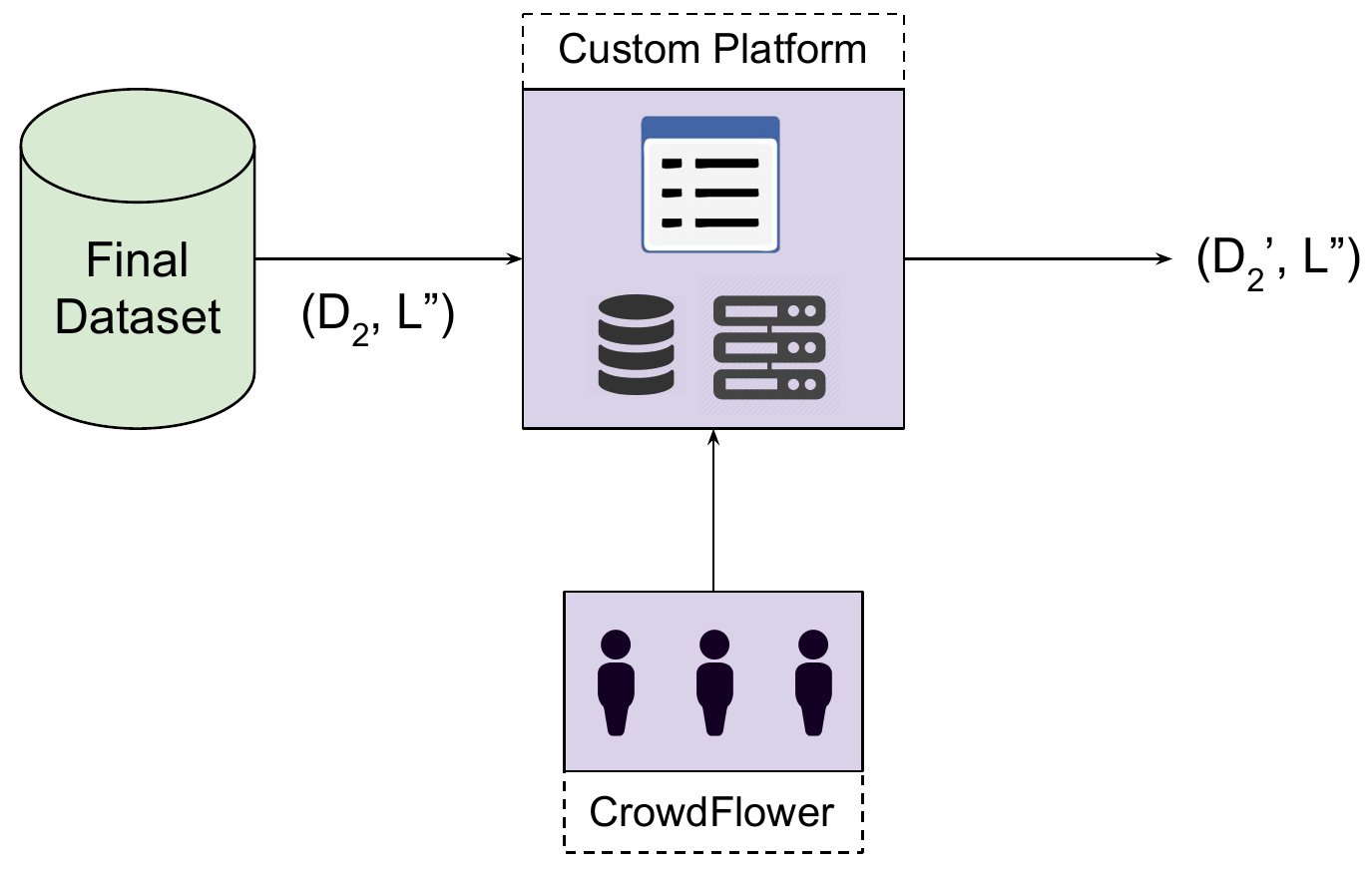} 
	\caption{Final Annotation Round (Step 3). A larger dataset $D2$, with the final label set $L''$ can be used for large scale annotation. A custom-built platform used allows for better control of the annotation flow, and reduces dependencies on CrowdFlower specific design limitations.}
	\label{fig:final}
	\vspace{-1em}
\end{figure}

The third step of the process is when we actually annotate the larger dataset ($D2$), using the previously established settings and labels ($L''$). As shown in Figure~\ref{fig:final}, we built our own custom platform to host the annotation task. To accommodate such a large-scale task, we also created a database schema to store the data and the results, and to calculate the statistics.

\section{Step 1: Data collection and sampling}\label{sec:data}

In this section, we present the data preparation procedure. We detail how we collect and filter the data, the preprocessing part, and finally the necessary sampling. The pipeline is shown in Figure~\ref{fig:datapipeline}.

\subsection{Collection}
The first step of the process is to collect a random set of tweets. To do so, we utilize the Twitter Stream API and we collect all the tweets provided by the API ($1\%$ of the entire traffic) during the period of 30th March 2017 - 9th April 2017, consisting of $32$ million tweets in total.

\subsection{Metadata Extraction}
To facilitate filtering and sampling (the next two steps), each tweet is enriched with metadata (Figure~\ref{fig:datapipeline}). First of all, from the tweet's content we extract the number of URLs, hashtags, mentions, emojis/smilies, and numerals. Furthermore, we tag retweets and mentions. Finally, we extract metadata from Twitter, such as the detected language, the account age, etc. Additionally, we apply sentiment analysis, such as polarity and subjectivity of the tweet, using the TextBlob Python library. Finally, we count the number of offensive terms found using two dictionaries (HateBase\footnote{\url{https://www.hatebase.org}} and an offensive words dictionary\footnote{\url{https://www.noswearing.com/dictionary}}).

\subsection{Filtering}
For the entire dataset, we apply some basic preprocessing in order to filter out tweets that should not be annotated. Firstly, we remove tweets that are considered spam. There are numerous  techniques for tweet spam detection \cite{dhingra2015content}, \cite{santos2014twitter}, \cite{wang2015making}, \cite{zhounetwork}, \cite{wang2010don}. Inspired by these, we apply filtering criteria for the elimination of such spam-related tweets. Furthermore, we only keep original tweets (i.e., drop retweets without new content), while also remove those that have small text content (e.g., only URLs, images, etc.). Finally, we remove any tweets that are not written in English, using Twitter's language detection.

\subsection{Boosted Sampling}
As shown in Figure~\ref{fig:datapipeline}, after we collect and clean the data, the next step is to create the final dataset that will be used in the various rounds. One major issue that needs to be addressed when considering such datasets is the class imbalance of the behavior under study. More particularly, in the case of abuse detection, even though inappropriate content is very frequent in Online Social Networks, it is still a minority compared to the tremendous amount of ``normal'' data produced. Therefore, when selecting the data to create a sample that will be annotated, it is necessary to ensure there will be plenty of inappropriate annotations to work with, otherwise the dataset is not very useful for the research community. Therefore, we follow a sampling procedure (BS) and inject the selected data in the randomly sampled (RS) ones.

For the boosted sample, we use the metadata extracted earlier. We choose tweets that, based on the sentiment analysis, show strong negative polarity ($< -0.7$)  and contain at least one offensive word. Finally, we create two datasets: $D_1$ is a sample of just $300$ tweets that is used for the exploratory analysis, and $D_2$  that contains $80K$ tweets that will be used for the final annotation.

\subsection{Datasets}
In total, we work with two datasets, $D_1$ and $D_2$, one for each step. In Table~\ref{tab:datasets} we present the datasets used per round with some extra information regarding the annotations. 

\begin{table}[t]
\setlength\tabcolsep{2pt}
	\centering
	\small
	\begin{tabular}{|lccccc|}
		\hline
		& Dataset & Tweets & Judg. & Labels & Sampling Percentage\\ \hline \hline
		\multicolumn{6}{|c|}{Step 2} \\ \hline
		R1 & $D_1$ & 300 & 5 & 7 & 33\% BS - 67\% RS \\
		R2 & Subset of $D_1$ & 88 & 10-20 & 7 & 92\% BS - 8\% RS \\
		VR & $D_1$ & 300 & 5 & 4 & 33\% BS - 67\% RS \\ \hline
		\multicolumn{6}{|c|}{Step 3} \\ \hline
		FR & $D_2$ & 80k & 5 & 4 & 12.5\% BS - 87.5\% RS \\  \hline
	\end{tabular}
	\caption{Datasets per Round}
	\label{tab:datasets}
	\vspace{-1em}
\end{table}

\section{Step 2: Exploratory Rounds} \label{sec:exploratory}

The goal of this step is to tune the crowdsourcing parameters on a smaller dataset, in order to quickly get some insights but minimize the cost while doing so.

We focused our exploration on identifying the most representative labels related to the types of abusive content. We begin with the most extensively used labels found in literature, and at each round we looked into the results and merge or remove labels that were frequently confused by the annotators. Furthermore, these rounds helped us to further filter spam and get a more representative boosted sampling.

We begin with a first round (R1) that includes 300 tweets and 5 judgments per tweet. We collect annotations and assess if the plurality of labels is confusing, how the spamming annotation works, etc. Afterwards, we continue to a second round (R2), where we focus only on the tweets that were  marked as inappropriate in the first round, but requesting a much larger number of annotations to assess better the overlap of used labels. Finally, we conclude with a third round (VR) to validate the selected labels and confirm annotation agreement, before moving on to the Step 3 and the large scale annotation (FR).

\subsection{Definitions}

Before starting, annotators are provided with definitions for each label which they have to acknowledge reading. The definitions are constructed based on all the descriptions we found in the related literature, as cited on each category, as well as Cambridge\footnote{http://dictionary.cambridge.org} and Black's Law\footnote{http://thelawdictionary.org} dictionaries. In total, the following definitions were displayed to the annotators:

\begin{itemize}
	
	\item {\bf Offensive Language}: Profanity, strongly impolite, rude or vulgar language expressed with fighting or hurtful words in order to insult a targeted individual or group. 
	\cite{chen2012detecting}, \cite{razavi2010offensive}.
	
	\item {\bf Abusive Language}: Any strongly impolite, rude or hurtful language using profanity, that can show a debasement of someone or something, or show intense emotion. 
	\cite{papegnies2017detection}, \cite{park2017one}, \cite{nobata2016abusive}.
	
	\item {\bf Hate Speech}: Language used to express hatred towards a targeted individual or group, or is intended to be derogatory, to humiliate, or to insult the members of the group, on the basis of attributes such as race, religion, ethnic origin, sexual orientation, disability, or gender. 
	\cite{davidson2017automated}, \cite{badjatiya2017deep}, \cite{warner2012detecting}, \cite{schmidt2017survey}, \cite{djuric2015hate}.
	
	\item {\bf Aggressive Behavior}: Overt, angry and often violent social interaction delivered via electronic means, with the intention of inflicting damage or other unpleasantness upon another individual or group of people, who perceive such acts as derogatory, harmful, or unwanted.
	\cite{chatzakou2017mean}, \cite{hosseinmardi2015analyzing}.
	
	\item {\bf Cyberbullying Behavior}: It's the use of force, threat, or coercion to abuse, embarrass, intimidate, or aggressively dominate others, using electronic forms of contact. It typically denotes repeated and hostile behavior performed by a group or an individual. 
	\cite{chatzakou2017mean}, \cite{hosseinmardi2015analyzing}, \cite{dinakar2011modeling}, \cite{riadi2017detection}, \cite{kansara2015framework}.
	
	\item {\bf Spam}: Posts consisted of related or unrelated advertising / marketing, selling products of adult nature, linking to malicious websites, phishing attempts and other kinds of unwanted information, usually executed repeatedly.
	
	\item {\bf Normal}: all tweets that do not fall in any of the prior categories.
	
\end{itemize}

Therefore, our starting set of labels $L$ is: 
\[
L=\{Offensive, Abusive, Hateful, Aggressive,
\]
\[
Cyberbullying, Spam, Normal\}
\]

\subsection{First Round}
On the first round, annotators were asked in a primary selection, to first classify tweets into three general categories: normal, spam, and inappropriate. In the case that inappropriate was selected, then a secondary panel offered them the five aforementioned inappropriate speech categories. This way, users could define more explicitly the type exhibited by the tweet. Furthermore, they had the option to suggest a new subcategory utilizing the ``other'' option and a text box. Finally, the participants were encouraged to select multiple subcategories whenever appropriate. The dataset described above was sent to CrowdFlower for annotation, asking for five judgments per tweet. 

\begin{figure}[t]
	\centering
	\includegraphics[width=\columnwidth]{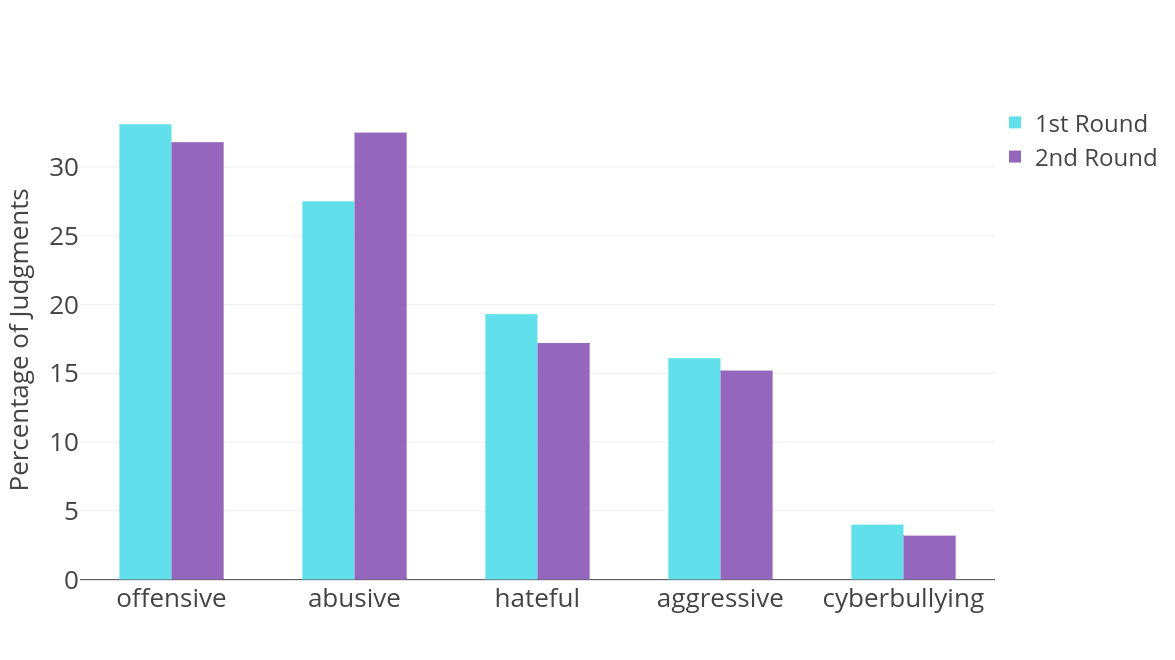}  
	\caption{Distributions of judgments per inappropriate label for the two exploratory rounds in Step 2.}
	\label{res:bardistributions}
	\vspace{-1em}
\end{figure}

In Figures~\ref{res:bardistributions} and~\ref{res:majority} we see (in blue color) some of the results of this round. Figure~\ref{res:bardistributions} shows the distributions of judgments per inappropriate label. It is important to note here that the percentages of both rounds take under consideration only the total amount of inappropriate labels, since these are the ones we want to observe. We notice that Offensive and Abusive are the most popular categories, followed by Hateful and Aggressive. Cyberbullying is rarely used. Normal and Spam are not presented on the figure, but are very frequently used, with a percentage of 53\% for Normal, and 15\% for Spam, overall.

In Figure~\ref{res:majority} we observe the agreement of the annotators, when there is majority voted, grouped in three majority categories, for convenience. The three categories are:
i) Overwhelming majority, when at least 80\% of the annotators agree,
ii) Strong Majority, when at least 50\% of the annotators agree and
iii) Simple majority, for the rest of the cases.
When two or more labels have equal number of judgments, the tweet is not included in any of the three categories, since it is not assigned with a majority vote. The results of the first round show a clear ``win'' of the Overwhelming category, which means that most participants agreed on their votes. On the other hand, most of the judgments of this round are Normal or Spam, therefore we can not be positive that the majority results refer to the inappropriate labels. The confusion becomes more clear when we run the second round, the results of which are presented below.

\subsection{Second Round}
Results presented on the first round provide some insights on the correlation among inappropriate speech categories. However, our confidence on these results is low, mainly because of the low amount of annotations per tweet. For this reason, we decide to proceed in a new annotation round. Here, we use only the tweets that were previously annotated as Inappropriate, with a high agreement score. In total, these tweets are 88 out of the initial 300. Each tweet was consequently annotated by at least 10 workers, but usually around 20. We kept the same setup regarding labels and instructions, as we want to be able to compare the two rounds afterwards.

\begin{figure}[t]
	\centering
	\includegraphics[width=\columnwidth]{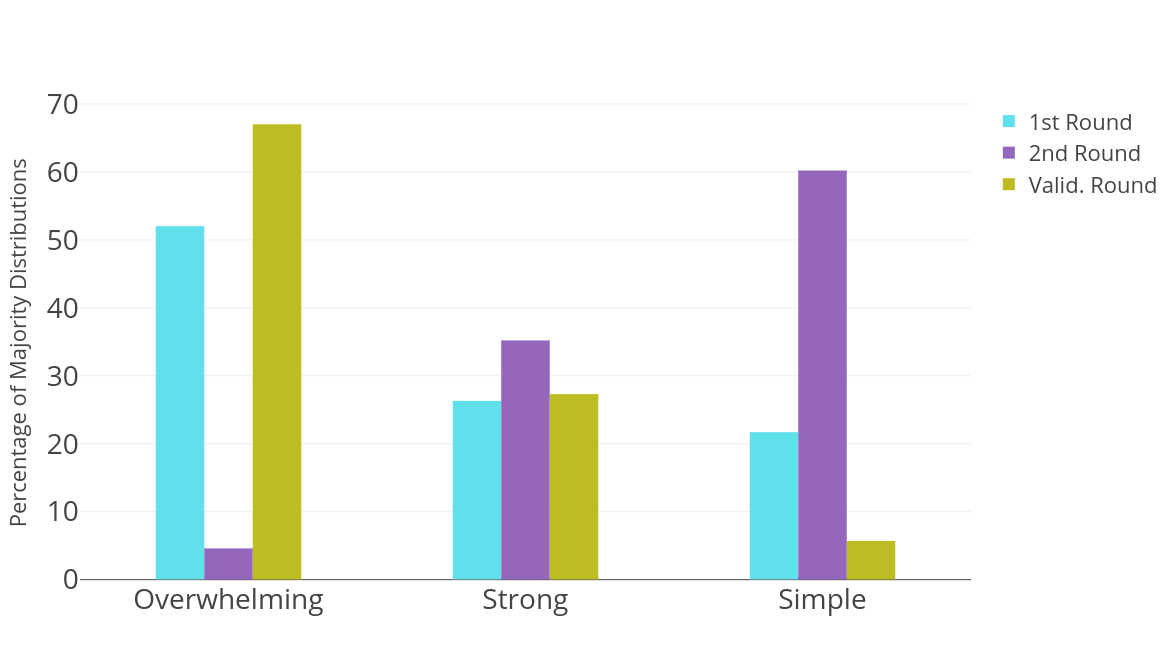}  
	\caption{Categories of majority distributions for all preliminary rounds in Step 2.}
	\label{res:majority}
	\vspace{-1em}
\end{figure}

The results of Figure~\ref{res:bardistributions} show a similar, yet not identical, distribution of the five labels. Offensive and Abusive are still the ``leading'' labels, although Abusive is slightly more popular in this round. Hateful and Aggressive follow again, in the same order. Finally, Cyberbullying is again very low. On the other hand, majorities in Figure~\ref{res:majority} have completely changed. As it appears, in most cases annotators disagree about the labels, and only very few have a high amount of agreement. This clearly shows that the task of choosing between our set of labels is not trivial.

\subsection{Comparison of the Exploratory Rounds}
In order to study how labels are related, we compare the results of the previous two rounds. More specifically, we calculate correlations of the various labels, measure their similarities and report on co-occurrences. In this section, we study these statistics and reach the final decision over which labels will be kept for the validation round.

We begin by measuring the correlation and similarities among the inappropriate category labels. Such correlations and similarities will allow us to measure how closely each label appears to have been selected with another label, given a set of tweets. We calculate correlations and their significance using Pearson, Spearman, and Kendall Tau Correlation Coefficients. The similarity between labels is measured using Cosine Similarity. For each pair of labels, we calculate their similarity vectors, in order to gain some insight on the correspondence of pairs in accordance with their ranking. That is, for each label, we construct a vector of votes, with each cell representing a tweet annotated. Then, we compute the similarity of these vectors, in all-pairwise fashion between the available labels. Similarly, we compute the correlation of labels using these vectors. In Table \ref{tab:correlations}, we present these results for both rounds, in order to compare.

\begin{table*}[t]
	\centering
	\small
	\begin{tabular}{|l|cccccc|c|}
		\hline
		& PCC & p-PCC & SCC & p-SCC & KTCC & p-KTCC & CosSim \\ \hline\hline
		\multicolumn{8}{|c|}{First Round (300 tweets)} \\ \hline
		Offensive - Abusive & 0.057672 & 0.4863 & 0.109460 & 0.1854 & 0.095695 & 0.0844 & 0.536908 \\
		Offensive - Hateful & -0.064017 & 0.4395 & -0.008290 & 0.9203 & -0.007995 & 0.8854 & 0.410749 \\
		\textbf{Offensive - Aggressive} & -0.122807 & 0.1370 & -0.124149 & 0.1327 & -0.110099 & \textbf{0.0471} & 0.348367 \\
		Offensive - Cyberbullying & -0.083271 & 0.3143 & -0.055111 & 0.5059 & -0.049595 & 0.3711 & 0.209020 \\
		Abusive - Hateful & -0.096501 & 0.2433 & -0.036050 & 0.6636 & -0.033111 & 0.5504 & 0.320653 \\
		\textbf{Abusive - Aggressive} & 0.195979 & \textbf{0.0170} & 0.324244 & \textbf{0.0001} & 0.285359 & \textbf{0.0000} & 0.478639 \\
		Abusive - Cyberbullying & 0.042049 & 0.6118 & 0.065070 & 0.4320 & 0.060229 & 0.2774 & 0.251285 \\
		Hateful - Aggressive & -0.042224 & 0.6104 & 0.024063 & 0.7716 & 0.020994 & 0.7050 & 0.279881 \\
		Hateful - Cyberbullying & -0.076778 & 0.3537 & -0.080666 & 0.3298 & -0.076305 & 0.1688 & 0.133986 \\
		\textbf{Aggressive - Cyberbullying} & -0.142836 & 0.0833 & -0.169740 & \textbf{0.0392} & -0.161572 & \textbf{0.0036} & 0.066667 \\ \hline
		\multicolumn{8}{|c|}{Second Round (88 tweets)} \\ \hline
		\textbf{Offensive - Abusive} & 0.322597 & \textbf{0.0022} & 0.408552 & \textbf{0.0001} & 0.294481 & \textbf{0.0000} & 0.741228 \\
		Offensive - Hateful & -0.076287 & 0.4799 & -0.130442 & 0.2258 & -0.095233 & 0.1889 & 0.544227 \\
		\textbf{Offensive - Aggressive} & 0.056567 & 0.6006 & 0.245213 & \textbf{0.0213} & 0.186104 & \textbf{0.0102} & 0.482113 \\
		\textbf{Offensive - Cyberbullying} & 0.230017 & \textbf{0.0311} & 0.191246 & 0.0743 & 0.157496 & \textbf{0.0298} & 0.497397 \\
		Abusive - Hateful & 0.126504 & 0.2402 & 0.118195 & 0.2727 & 0.079851 & 0.2706 & 0.619584\\
		\textbf{Abusive - Aggressive} & 0.011576 & 0.9148 & 0.270948 & \textbf{0.0107} & 0.199341 & \textbf{0.0060} & 0.451047\\
		\textbf{Abusive - Cyberbullying} & 0.243139 & \textbf{0.0225} & 0.241344 & \textbf{0.0235} & 0.202128 & \textbf{0.0053} & 0.501380\\
		Hateful - Aggressive & -0.072054 & 0.5047 & -0.039991 & 0.7114 & -0.030565 & 0.6733 & 0.374166\\
		Hateful - Cyberbullying & 0.013788 & 0.8985 & 0.003095 & 0.9772 & 0.003917 & 0.9569 & 0.350230 \\
		Aggressive - Cyberbullying & -0.001821 & 0.9866 & 0.125888 & 0.2425 & 0.109380 & 0.1313 & 0.268009 \\ \hline
	\end{tabular}
	\caption{Correlation Coefficients, p-values and Cosine Similarity values for each pair of inappropriate labels in the two Exploratory Rounds.}
	\label{tab:correlations}
	\vspace{-1em}
\end{table*}

On the first round, all three correlation coefficient metrics show low correlation between most of the labels. The only correlation that seems statistically significant in all cases is between Abusive and Aggressive ($p$$<$$0.05$). Moreover, Aggressive and Cyberbullying seem to be somewhat correlated, but the significance is not consistent. Finally, Kendall Tau also shows a significant relationship between Offensive and Aggressive, which does not appear in the other two cases. 
Other combinations do not exhibit important correlations.

When it comes to the second round, there are generally low correlations with no statistical significance between the labels, with some exceptions. Offensive and Abusive are correlated in statistically significant fashion ($p<0.05$), and this is consistent across all metrics. Furthermore, in most of the cases, both are also significantly correlated to Cyberbullying. Spearman and Kendall also show some correlation between Offensive and Abusive with Aggressive. Hateful, never seems to be correlated with the rest of the labels.

Regarding the cosine similarities, we see in Table~\ref{tab:correlations} that in the first round, the values of similarities are not very high. Nevertheless, the most highly similar pairs are Offensive and Aggressive with Abusive. 
Alternatively, during the second round, the values of the similarities are much higher than before. Again the most similar pair of labels is Offensive - Abusive, followed by Abusive - Hateful. We notice in general, in this annotation round, that hateful seems to be more related than it was on the previous round, but the correlation results still don't indicate a strong correspondence.

To support the previous results, we also calculate the co-occurrences of the various labels for each one of the three majority agreement groups (Overwhelming, Strong and Simple). Due to space limitations, we can not fully present the co-occurrences results here. However, we briefly state what we observed and how they support our final learnings. The results show that users seem to be very confused about selecting a label, resulting in low levels of agreement for most of the inappropriate tweets. On the second annotation round, for example, Abusive seems to be used a lot of times and is often the majority label, but it's always confused with many of the other labels (especially Offensive). Offensive is also confused with Abusive and Aggressive, and frequently also with Hateful. Finally, Cyberbullying never becomes a majority-label. As expected, this becomes more intense with harder to categorize tweets, i.e., tweets that most annotators disagreed in their judgments. 

\subsubsection{Insights learned:}

The results of the first two annotation rounds allowed us to draw the following conclusions regarding the use of the five inappropriate speech labels:
\begin{enumerate}
	\item We can form three groups of labels according to their popularity: Abusive and Offensive are the two most popular labels, Hateful and Aggressive are somewhat popular, and Cyberbullying is rarely used.
	\item Cyberbullying can be safely eliminated from the list of inappropriate labels, mainly due to the very few times it was selected on both the annotation rounds. This decision, though, is also supported by the very nature of Cyberbullying, which according to its definition should be repetitive. Yet, in our case we have no sense of time or repetition, since we work with individual tweets.
	\item Abusive, Offensive and Aggressive seem to be significantly correlated, highly coexisting in the annotations and very similar (according to the similarity results). Abusive is the most popular among the three and the most central (i.e. the other two labels are much more related with this than with each other).
	\item While Hateful is frequently coexisting with other labels, indicating a confusion among users in the use of this label, it does not appear to be significantly correlated with any other label. This is also supported by the definition of Hateful, since there is a well-defined description of the target groups of this category, compared to the rest.
\end{enumerate}

\subsection{Validation Round}

Given the above insights, we proceed with extra validation rounds, before the large-scale annotation. To do so, we first remove Cyberbullying (due to point 1). Then, utilizing the insights from point 2, we merge Abusive, Offensive and Aggressive into a single category. To choose which label to use, we run one annotation campaign for each keyword, and we achieved similar results. Therefore, we kept Abusive as the keyword for this category of tweets. Finally, we decided to keep Hateful separately, as explained in point 3. Thus, in these validation rounds, users were presented with 4 labels:
\[
L'=\{Abusive, Hateful, Normal, Spam\}
\]
The dataset used is again $D_1$ (containing 300 tweets) and we required again 5 judgments per tweet.

\begin{figure}[t]
	\centering
	\includegraphics[width=\columnwidth]{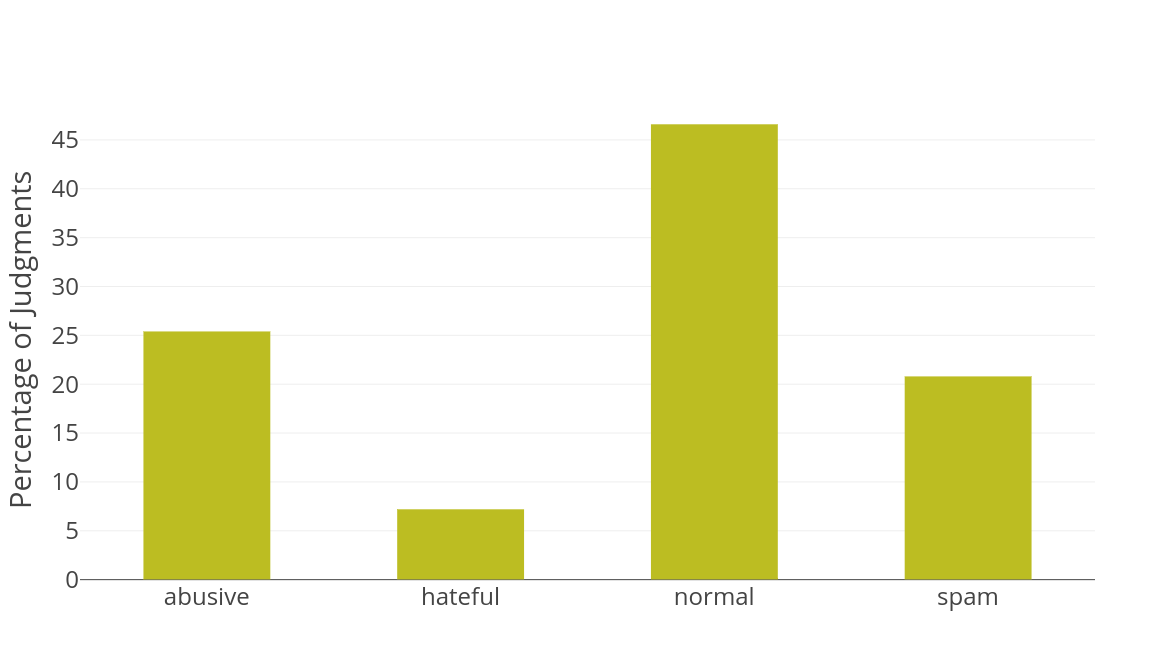}  
	\caption{Distributions of judgments per inappropriate label for the validation round.}
	\label{res:valroundjudge}
	\vspace{-1em}
\end{figure}

We begin with Figure~\ref{res:valroundjudge}, where we notice the distribution of judgments towards labels. We see that the final inappropriate labels are much more frequently used now than before, something that was expected as we merged most of them. Hateful is still not as frequent as Abusive, but it appears in almost 7\% of the judgments, therefore can not be eliminated. In the agreement graph of Figure~\ref{res:majority}, we clearly notice a vast improvement from the previous rounds. Almost 70\% of the tweets reach an Overwhelming Agreement (more than 3 out of 5 annotators agree), while annotators disagree highly only in very few tweets. This result is of course highly connected with the fact that we presented a simpler task for the workers to complete.

The correlations coefficient and similarities table (Table~\ref{tab:correlationsval}) also depict some consistent results. More specifically, we have statistically significant negative correlations when observing the interactions of Abusive with all other labels (i.e., the appearance of the Abusive label is negatively correlated with the appearance of the other labels) and no important correlation in any other combination of labels. In the last column, we see the cosine similarities between the pairs. We notice that the vectors are much less similar, while also Abusive is clearly different than Normal and Spam and not very similar with Hateful either. Finally, from the co-occurrences results we saw that Abusive and Hateful are still sometimes confused (showing that our task is still not trivial), but this is not very frequent across tweets.

\begin{table*}[t]
	\centering
	\small
	\begin{tabular}{|l|cccccc|c|}
		\hline
		& PCC & p-PCC & SCC & p-SCC & KTCC & p-KTCC & CosSim \\ \hline
		Abusive - Hateful & -0.378388 & 0.0000 & -0.436322 & 0.0000 & -0.375022 & 0.0000 & 0.382150 \\
		Abusive - Normal & -0.839857 & 0.0000 & -0.843253 & 0.0000 & -0.738567 & 0.0000 & 0.195936 \\
		Abusive - Spam & -0.314023 & 0.0001 & -0.350580 & 0.0000 & -0.304493 & 0.0000 & 0.136757 \\
		Hateful - Normal & -0.068685 & 0.4101 & 0.024373 & 0.7703 & 0.014200 & 0.7992 & 0.406726 \\
		Hateful - Spam & -0.072026 & 0.3876 & -0.027309 & 0.7435 & -0.025878 & 0.6430 & 0.194662 \\
		Normal - Spam & 0.028358 & 0.7340 & 0.130268 & 0.1171 & 0.111400 & 0.0460 & 0.267608\\ \hline
	\end{tabular}
	\caption{Correlation Coefficients, p-values and Cosine Similarity values for the labels of Validation Round.}
	\label{tab:correlationsval}
\end{table*}

\section{Step 3: Large Scale Annotation} \label{sec:annotation}

Based on the decisions drawn upon the previous results, we launched the final, large-scale annotation task of 80k tweets, with 5 judgments per tweet. The setup for this round is kept the same with the last validation round, since it's already tested. Thus, the final labels we decided upon, as analyzed earlier, are:
\[
L''=\{Abusive, Hateful, Normal, Spam\}
\]

\subsection{Annotator Profiles}

In order for CrowdFlower workers to participate in the annotation task, we first require some basic demographic information, such as gender, age, annual income, level of education and nationality. This is to have a better understanding of the annotators' profiles. Here, we analyze these demographics. First of all, we start with gender. Almost two thirds of the participants are male (66.6\%) and one third is female (33.4\%). Moreover, even though we provided an ``other'' option, no workers selected it. Regarding their educational level, most of them have a Bachelor Degree (48.4\%), followed by Secondary Education (29.9\%) and Master's Degree (20.1\%), while very few have a PhD (1.6\%).

The age of the participants ranges from 18 to over 87; 32.6\% are between 18 and 24, 29.1\% are 25-31, 18.2\% are 32-38 and the remainder above 39 years old. More than half of them claim to have an income level below \euro10k (57.7\%), 14.4\% are between \euro10k and \euro20k and the rest spread across \euro20k and \euro100k. Finally, their nationalities vary a lot, coming from 114 different countries in total. However, by far the most frequent country of origin is Venezuela (48\%), followed by USA (9.8\%), Egypt (6.3\%) and India (4\%). Overall, the annotators from the top 10 countries contribute 81.2\% of all annotations.

\subsection{Results}

\begin{figure}[t]
	\centering
	\includegraphics[width=\columnwidth]{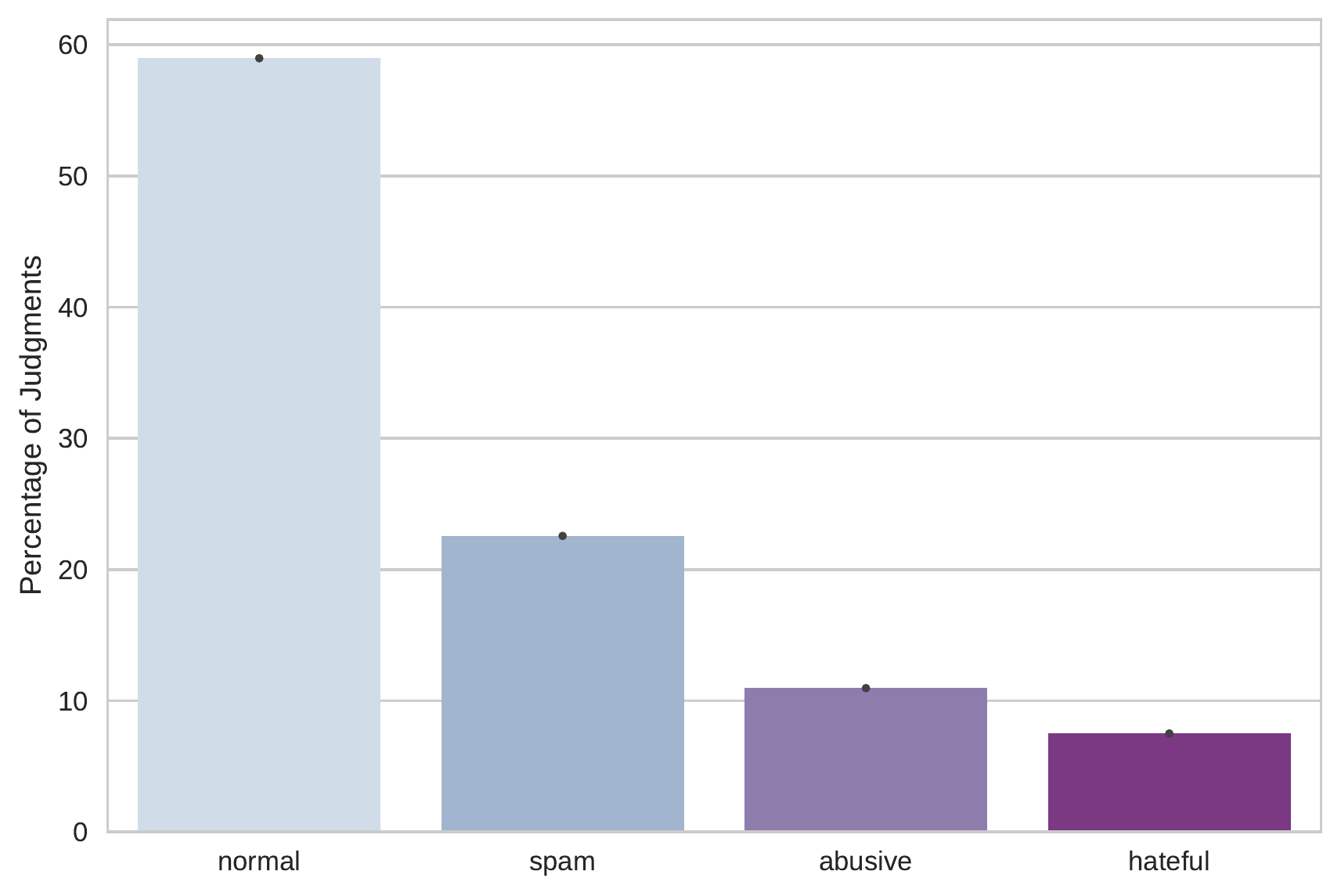}  
	\caption{Label distributions of the large-scale annotated dataset.}
	\label{res:finalroundlabeldistributionslarge}
	\vspace{-1em}
\end{figure}

The distributions of the final judgments (Figure~\ref{res:finalroundlabeldistributionslarge}) are quite similar with the preliminary rounds. The two inappropriate labels cover almost 20\% of all judgments. Abusive is again more popular (11\%) compared to hateful (7.5\%), and normal is still the most popular label (59\%). However, the distribution of spam is somewhat different, since now it is more popular than the inappropriate labels (22.5\%). This change is due to the imbalanced sampled subsets (the portion of random samples is much higher than the boosted samples).

\begin{figure}[t]
	\centering
	\includegraphics[width=\columnwidth]{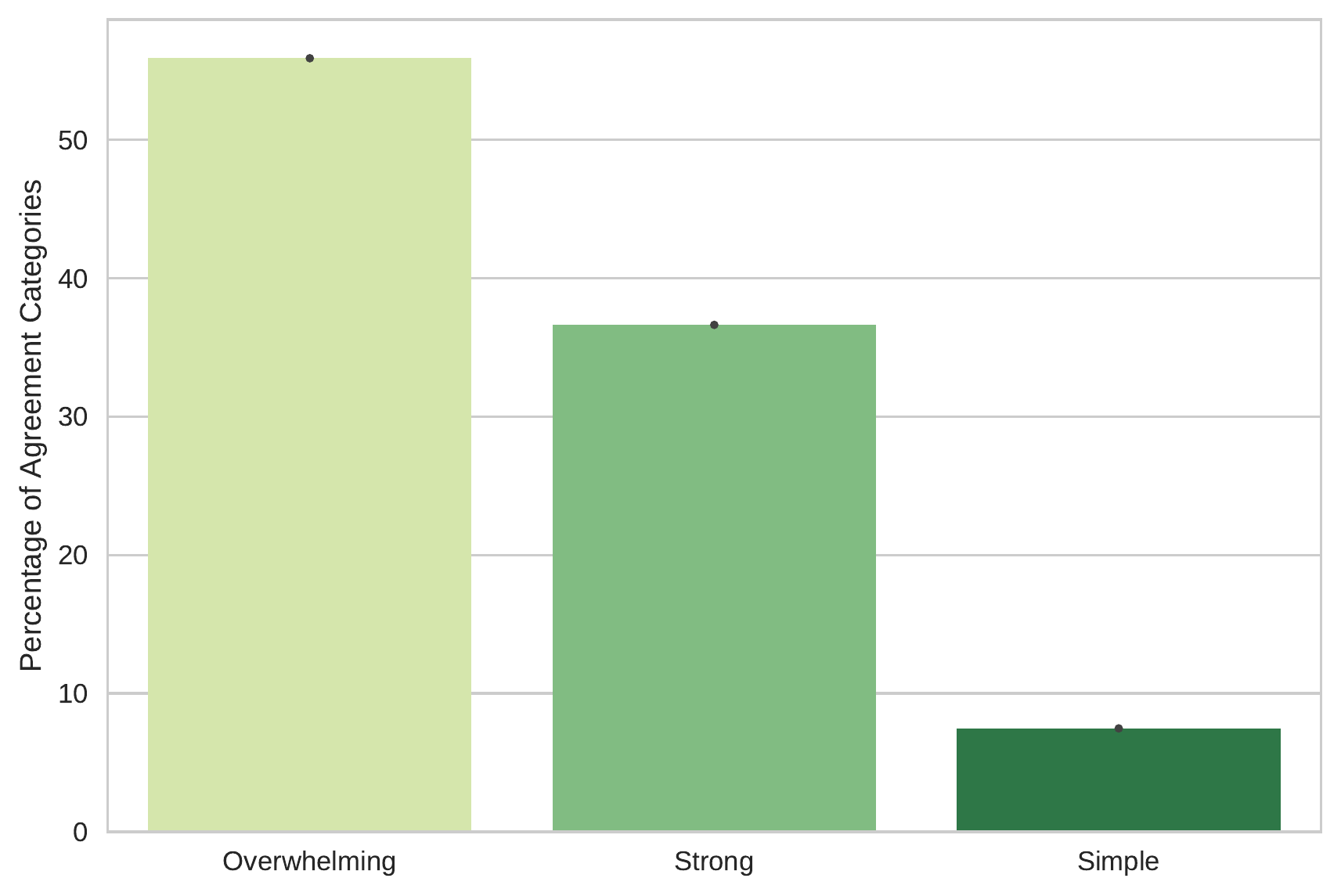}  
	\caption{Distributions of categorized agreement for the large-scale annotated dataset}
	\label{res:majoritieslarge}
	\vspace{-1em}
\end{figure}

Regarding the three categories of agreement scores (Overwhelming, Strong and Simple) shown in Figure~\ref{res:majoritieslarge}, we observe that more than half of the tweets ($\sim$55.9\%) achieve an overwhelming agreement, which means that at least 4 out of 5 annotators agreed on the label. The remaining $\sim$36.6\% of tweets reach an agreement of more than 3 out of 5 votes and only very few ($\sim$7.5\%) achieve majority with only two annotators. In general, these majorities show a very strong agreement between the annotators, therefore we consider our results robust. \

\begin{figure}[t]
	\centering
	\includegraphics[width=\columnwidth]{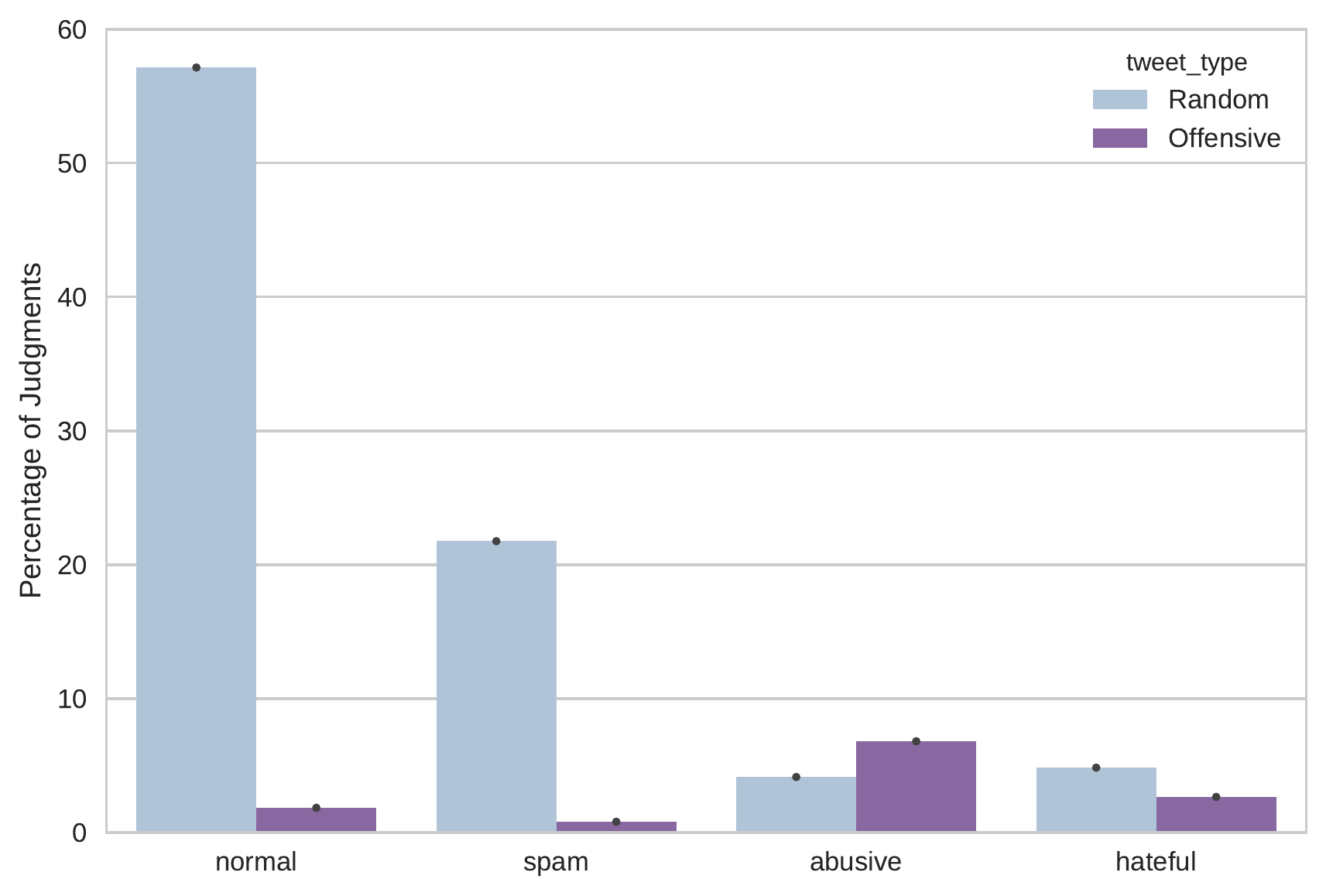}  
	\caption{Label distributions for Random vs Boosted-Sampled categories.}
	\label{res:randomsampled}
	\vspace{-1em}
\end{figure}

Finally, we compare the two sampled categories (Random vs Boosted sample) in Figure~\ref{res:randomsampled}. We clearly see that, as expected, the boosted sample is by far richer in Abusive and Hateful content than the random sample. The overall percentage of inappropriate tweets is not very high, due to the fact that the samples are imbalanced (as mentioned earlier the final dataset consisted of two subset, the random, and the boosted, consisting of 70k and 10k entries respectively).

For the sake of deeper understanding, we also compare a balanced set of 10k random with the 10k boosted. From this comparison, we get a much higher percentage of inappropriate tweets, that sum to a total of almost half the annotations. In fact, there are almost as many Abusive tweets in boosted, as Normal in random ($\sim$35\% in both cases), and a very low total of $\sim$4\% Abusive and Hateful tweets in the random sample. This means that our decision to use a boosted sample in our methodology proved crucial: the amount of inappropriate tweets would have been too low to produce any important results, without the boosted sample.

\section{Conclusion}\label{sec:discussion}

In this work, we provided a methodology for annotating a large-scale dataset of inappropriate speech and the resulting labeled dataset. This annotation focused on various facets of abusive or hateful language in Twitter. We selected these two types, out of several inappropriate speech categories, based on an empirical analysis of the relationships between the corresponding labels. More specifically, we selected the most popularly used types of inappropriate speech in literature, and conduct a series of annotation rounds to understand how crowdworkers use these labels.

We analyzed these annotations in terms of correlations and similarities between the labels, and calculated their co-occurrences. After statistical analysis of these similarities between labels, we merged some and eliminated others, to conclude to the most representative set. In this case, it was Abusive and Hateful - and eliminated some less useful such as Cyberbullying. When we obtained the final structure of the annotation task, we annotated the large-scale dataset.

With this present work, we make available 1) our followed methodology, 2) our code used for the custom-built platform for annotation, and 3) our final annotated dataset. We hope these three items will be useful to the research community of either performing annotations, or building machine learning models on top of such datasets. As a future work, we plan on enriching the dataset with more boosted data, since, as we showed, they carry most of the valuable information about inappropriate speech. The updated versions of the dataset will be maintained here: \url{https://github.com/ENCASEH2020/hatespeech-twitter}.

\section{Acknowledgements}\label{sec:acknowledgements}

The authors acknowledge research funding from the European Union’s Horizon 2020 research and innovation programme under the Marie Skłodowska-Curie Grant Agreement No 691025.

\small
\bibliography{bibfile}
\bibliographystyle{aaai}
\end{document}